\documentclass[%
 reprint,
 amsmath,amssymb,
 aps,
]{revtex4-2}

\usepackage[pdfborder={0 0 0}]{hyperref}
\usepackage{graphicx}
\usepackage{dcolumn}
\usepackage{bm}
\usepackage{xcolor}
\usepackage{soul}

\begin{document}

\preprint{APS/123-QED}

\title{Tunneling amplifies chirality-induced spin selectivity and explains its \\ current-direction invariance}

\author{F.~Baranov}
\email{fedor.baranov@fu-berlin.de}
\author{V.~Gali}
\affiliation{Dahlem Center for Complex Quantum Systems, Halle-Berlin-Regensburg Cluster of Excellence CCE, and Fachbereich Physik, Freie Universit\"at Berlin, Arnimallee 14, 14195 Berlin}

\author{G.~Lemut}
\author{M.~Breitkreiz}
\email{breitkr@physik.fu-berlin.de}
\affiliation{Dahlem Center for Complex Quantum Systems, Halle-Berlin-Regensburg Cluster of Excellence CCE, and Fachbereich Physik, Freie Universit\"at Berlin, Arnimallee 14, 14195 Berlin}

\date{\today}

\begin{abstract}
We propose a minimal model for chirality-induced spin selectivity (CISS) in dc transport through insulating chiral molecules, based on quantum tunneling and interaction-induced spin splitting. As a concrete realization of the latter, we consider a weak Zeeman interaction of the particle spin with the current-induced magnetic field, recently shown to occur in helical molecules. We show that quantum tunneling, combined with dissipation, amplifies the effect, so that even such a small spin-dependent perturbation can yield spin polarizations on the order of 100\% across a wide range of applied bias voltages. Furthermore, our tunneling scenario naturally reproduces the characteristic CISS symmetry of the current-voltage dependence---namely, the invariance of the spin-polarization sign under reversal of the current direction---while fully respecting Onsager's reciprocity relations.
\end{abstract}

\maketitle

\section{\label{sec:level1}Introduction\protect}

The chirality-induced spin selectivity (CISS) \cite{ ciss1999,Evers2022,Bloom2024,Yan_2024} has attracted considerable attention in recent years due to the unexpectedly large spin polarization (SP) observed in the transport behavior of chiral molecules. Key aspects of the transport mechanism, including the role of non-equilibrium conditions and the origin of spin splitting, are not yet fully understood. Some of the transport experiments are two-terminal magnetoconductance measurements, where a current is applied between a magnetic substrate and a non-magnetic scanning-tunneling-microscopy tip via long ($\sim 10$ nm) chiral molecules that are placed on the substrate. The magnetoconductance is the change in magnitude of the current $I_+-I_-$ at the same bias voltage between the electrodes upon switching the magnetization direction of the substrate. Since reversing the magnetization direction flips the spin of majority charge carriers in the substrate, the relative difference $(I_+-I_-)/(I_++I_-)$ is interpreted as spin polarization (SP) of the current. This SP is experimentally observed to reach values on the order of $100\%$, which increase with molecular length \cite{length1, length2}. The sign of the SP stays invariant upon reversal of the current direction but is opposite for opposite chiralities of the molecules. In this work we focus on this two-terminal dc transport phenomenon, which in the following we will refer to as just CISS for convenience. 

A central difficulty in understanding CISS can be expressed in terms of two fundamental challenges. First and on the quantitative level, it is generally difficult to obtain a large SP given the weak intrinsic spin-orbit coupling, which is the necessary ingredient linking the spin with the motion. Second and on the qualitative level, it remains nontrivial to explain the observed symmetry property of CISS \cite{Yang2020,Zhao2025}, namely the invariance of SP  under reversal of the current direction, and to reconcile it with Onsager's reciprocity relations.  A number of theoretical models invoking spin-dependent dissipation \cite{Matityahu2016,Volosniev2021}, electron-phonon  \cite{Wu2021,Fransson2020, Zhang2020,Kato2022,Klein2023} and electron-electron interactions \cite{ fransson2, Fransson2019a, Fransson2021a,Li2020, Chiesa2024, Herbrych2026} provided insights into possible  enhancement mechanisms of SP from the interplay of coherent and incoherent dynamics during electron transfer. While these models  reproduce some aspects of experimental observations, they fall short in others. 
A further complication is that a large SP is observed in setups typically featuring a film of molecules on top of the substrate, for which it is unclear how well the employed theoretical models correspond to the actual experimental conditions. Experiments with higher controllability in the form of single-molecule break junctions \cite{Li2025}, where the junction consists of a single chiral molecule rigidly coupled to two electrodes, did not detect SP.

This work focuses on two aspects that have received less attention from the theory side so far: the symmetry property of the CISS effect and the fact that CISS is experimentally observed in insulating molecules \cite{Yan_2024, Zhao2025}, while many theoretical calculations assume the molecules to be conducting. The insulating feature of systems relevant for the CISS effect is that their current-voltage characteristics with currents $I\sim$ nA at voltages $V_\mathrm{b}\sim 1$ V indicate that the molecules constitute extended potential barriers and the charge current is the result of quantum tunneling \cite{Nitzan2003}.
In the tunneling framework, in which the electronic probability amplitude decays inside the molecule, small perturbations can become strongly amplified in the transmission probability if they modify the decay length \cite{Pazy_2000, Burin2001}. We show that, in contrast to single-particle models, spin-dependent perturbations stemming from correlations can render the decay length spin-dependent. We specifically consider the effective Zeeman splitting  arising from a current-induced magnetic field, which has been recently shown to  reach the order of one Tesla \cite{fourcurrent,Bro2025}. 
While the resulting Zeeman splitting of $\sim 10^{-4}$ eV is still too weak to produce substantial SP in the conducting regime, we show that in the tunneling regime this spin-dependent modification of the tunneling barrier can produce SPs of order $100\%$ at low temperatures, which increase with molecular length. We further demonstrate that while the large polarization occurs in a very narrow energy range near the barrier edge, taking into account generic (spin-independent) dissipation at the interface between the tip-electrode and the molecule leads to an extended voltage range over which the large SP can be observed. 

Moreover, in such a tunneling scenario, at finite voltage bias, a reversal of the applied voltage switches between whether the tunneling goes via the occupied molecular energy level below the gap or the unoccupied level above it. Together with the flipped current, induced magnetic field, and thus flipped effective Zeeman splitting, this leads to voltage-sign invariance of the SP and thus naturally explains the observed symmetry of the CISS effect. Thereby, Onsager reciprocity relations do not apply, since the SP in our model only occurs in the non-linear response regime.

The remainder of this paper is organized as follows: in Section~\ref{single particle}
we discuss why the single-particle description of the effect is not sufficient for spin  splitting. In Section~\ref{2} we develop a formalism that describes transport in the presence of a self-induced magnetic field as a source of non-zero SP. In Section~\ref{symmetry} we address the resulting symmetry of the SP with respect to bias-voltage sign. In the following Sections \ref{zeroT} and \ref{5} we show and analyze numerical results for the SP at zero and finite temperature, respectively. We conclude with a discussion of our results in Section~\ref{summary}. 

\section{Insufficiency of single-particle description}
\label{single particle}

Before describing our model, we first show that the amplification of weak spin splitting in the tunneling regime requires going beyond a single-particle model, which explains why no large SP in single-particle tunneling descriptions of CISS has been obtained previously \cite{Mena2024}. The reason is rooted in Kramers' degeneracy of  tunneling probabilities in time-reversal invariant systems \cite{Macedo1992,Bardarson2008}, which we here reproduce within the tunneling context. Here, the spin of the tunneling particle couples only to its own momentum along the tunneling direction as a result of spin-orbit coupling, which we do not need to specify here. A pair of electronic modes, $\varepsilon_\pm(p)$, corresponding to Kramers' partners and given as a function of momentum $p$, takes the form
\begin{equation}
    \varepsilon_\pm(p) = e(p) \pm o(p)
    \label{kramer}
\end{equation}
where $e(p)$ and $o(p)$ are even and odd real functions of momentum, respectively. The form of Eq.\ \eqref{kramer} is due to time-reversal symmetry, which dictates that for each energy level there is the Kramers' partner level of opposite momentum, $\varepsilon_+(p)=\varepsilon_-(-p)$. Note that the spin polarizations (SPs) of a Kramers' pair of states are opposite since those states are related by time-reversal.

The decay length of a wavefunction at an energy $E$ below/above the lower/upper bound of the band $\varepsilon_\pm(p)$ is given by the inverse of the imaginary part of the momentum solutions of $E=\varepsilon_\pm(p)$.
If $p$ is a solution of $E=\varepsilon_+(p)$, then the complex conjugate $p^\star$ is also a solution, which can be seen by taking the complex conjugate of \eqref{kramer} leading to $E=\varepsilon_+(p^\ast)$. Hence, the imaginary parts of the solution pair $(p, p^\star)$ are  $(\text{Im}\: p, -\text{Im}  \:p)$. For the other mode, the solutions of $E=\varepsilon_-(p)$ are $-p$ and $-p^\star$, which follows from the even/odd character of the functions $e(p)$ and $o(p)$. The imaginary parts of this solution pair $(\text{Im}\: p, -\text{Im}  \:p)$ are, however, the same. Hence, the decay lengths of the two Kramers-partner states are equal, which means that for any spin-polarized decaying wavefunction inside the molecule there is an oppositely spin-polarized wavefunction with exactly the same decay length. This implies identical tunneling probabilities for opposite spins.

Interactions, which are not captured in the picture above, have the potential to circumvent this restriction. One obvious way is an interaction-induced spontaneous breaking of time-reversal symmetry. Another way is the interaction with particles under non-equilibrium conditions, which can be seen as a picture in which time-reversal symmetry is broken by the transport itself. The latter way is considered in the following, exemplified by the recently proposed spin splitting due to current-induced magnetic fields.

\section{Spin polarization from a self-induced magnetic field \label{2}}

When charge carriers tunnel through a helical molecule, the charge current density can generate a local magnetic field $\mathbf{B}$ at the position of the particle according to the Biot-Savart law. Recent studies \cite{fourcurrent, Bro2025} have shown that in molecules with helical structure, magnetic fields along the helical axis — induced by charge currents flowing through the molecule — are odd in the chirality and can reach values of up to $\sim1\,\text{T}$. 
This magnetic field couples to the electronic spin $\bm{S}$ via the relativistic Zeeman coupling $\frac{e}{mc} \boldsymbol{S} \cdot \boldsymbol{B}$, where $e$ and $m$ are the electron charge and mass, respectively, and $c$ is the speed of light. We here adopt a qualitative description, writing the Zeeman energy as
\begin{equation}
    H_\mathrm{Z} = -\sigma\,\chi\,\alpha (I), 
\end{equation}
where $\sigma=\pm$ is the sign of the spin projection along the  molecular helical axis, $\chi=\pm$ is the chirality, and $\alpha(I)=-\alpha(-I)$ parametrizes the Zeeman splitting, which is an odd function of the total current $I$ through the molecule. For numerical calculations below we model this dependence as 
\begin{equation}
    \alpha(I) = \alpha_0 \, I /I_0,
\end{equation}
where $I_0$ is taken to be the maximal current reached at voltages corresponding to the molecular energy level.  
According to Ref.\ \cite{fourcurrent}, the order of magnitude in long helical structures can reach $\alpha_0\sim10^{-4}$ eV.

We model the transmission through the molecule as an electronic 1D scattering problem, in which the molecule constitutes a homogeneous potential barrier of length $L$. Denoting the energy of the Zeeman-split molecular orbital as $V_\sigma = V_0 - \sigma\, \chi\, \alpha(I)$, the transmission probability for spin projection $\sigma$ at energy $E$ reads \cite{cohentannoudji1977quantum}
\begin{equation}
\label{T}
T_\sigma(E) = \left[ 1 + \frac{V_\sigma^2 \sinh^2\left( L \sqrt{\frac{2m(V_\sigma - E)}{\hbar^2}} \right)}{4E(V_\sigma - E)} \right]^{-1}.
\end{equation}
Via analytic continuation, this expression is valid also for energies above the barrier, in which case the argument of the hyperbolic sine is imaginary.

Before discussing the spin polarization of the current, it is instructive to first consider the relative transmission difference for opposite spin projections, as this clarifies its scaling with system parameters,
\begin{equation}
\label{Tinit}
T_{\mathrm{split}}=\frac{T_+ - T_-}{T_+ + T_-}.
\end{equation}
A simple, approximate but qualitatively similar expression for $T_{\mathrm{split}}$ in the tunneling regime $E<V_0-\alpha(I)$ can be obtained by using the Wentzel-Kramers-Brillouin (WKB) approximation 
\begin{equation}
    T_\sigma^{\mathrm{WKB}} =\exp\left(-2L\sqrt{2m(V_0-\sigma \chi \alpha(I) - E)}/\hbar \right),
\end{equation}
which leads to
$T_{\mathrm{split}} \approx \chi \tanh \big[ L\kappa_0\big(\sqrt{1+\epsilon} -\sqrt{1-\epsilon} \big)\big] $, where $\kappa_0 = \sqrt{2m(V_0-E)}/\hbar $ and $\epsilon =\alpha/(V_0-E)$. For a qualitative parameter dependence we can simplify $\Big(\sqrt{1+\epsilon} -\sqrt{1-\epsilon} \Big)\sim\epsilon$ to obtain
\begin{equation}
    T_{\mathrm{split}} \approx \chi \tanh \Bigg(\frac{L}{\ell_\alpha}\sqrt{\frac{\alpha(I)}{V_0-E}}\Bigg), \ \ \  \ \ell_\alpha\equiv\frac{\hbar}{\sqrt{2m\alpha(I)}},
    \label{eq:tsplit}
\end{equation}
where $\ell_\alpha$ is the characteristic mean free path associated with the action of the Zeeman interaction.  The maximal splitting is obtained when the energy reaches the lower energy barrier, $E=V_0-\alpha(I)$, in which case $ T^\mathrm{max}_{\mathrm{split}} = \chi\tanh(L/\ell_\alpha)$. This shows that strongly different transmissions of opposite spins can be obtained even for weak Zeeman splitting in long molecules when the length of the molecule is on the same order as the mean free path associated with the Zeeman splitting. For induced magnetic fields of order $1$ T  \cite{fourcurrent}, the mean free path is $\ell_\alpha\sim 10$ nm for a particle with the bare electron mass. Note that the effective mass of the particle can be larger, for example due to interaction with vibrations \cite{Klein2023}, which would further reduce $\ell_\alpha$; but already $\ell_\alpha\sim 10$ nm is on the order of the typical molecule length in relevant experiments.

The SP of the transmitted charge current we quantify in the standard way by calculating the relative difference of the transmitted current for opposite spins,
\begin{equation}\label{eq:SP}
    P = \frac{|I_+ - I_-|}{|I_+ + I_-|}.
\end{equation}
The current for each spin is given by the Landauer-B\"uttiker formula
\begin{equation}
\label{current}
    I_\sigma= \frac{e}{h} \int T_{\sigma} (E) \cdot [f_L(E) - f_R(E)] dE,
\end{equation} 
where $f_{L/R}(E)$ are the Fermi-Dirac distribution functions in the left and right leads, whose bias-voltage-dependence will be specified in the next section. 
At zero temperature, Eq.~\eqref{current} simplifies to an integral of $T_\sigma (E)$ over the energy window set by the applied voltage. 
Note that the transmission depends on the total current $I=I_++I_-$ through $\alpha(I)$, so that \eqref{current} needs to be solved self-consistently. 
\section{Symmetry of the I-V curve}
\label{symmetry}
\begin{figure}
\includegraphics[width=\columnwidth]{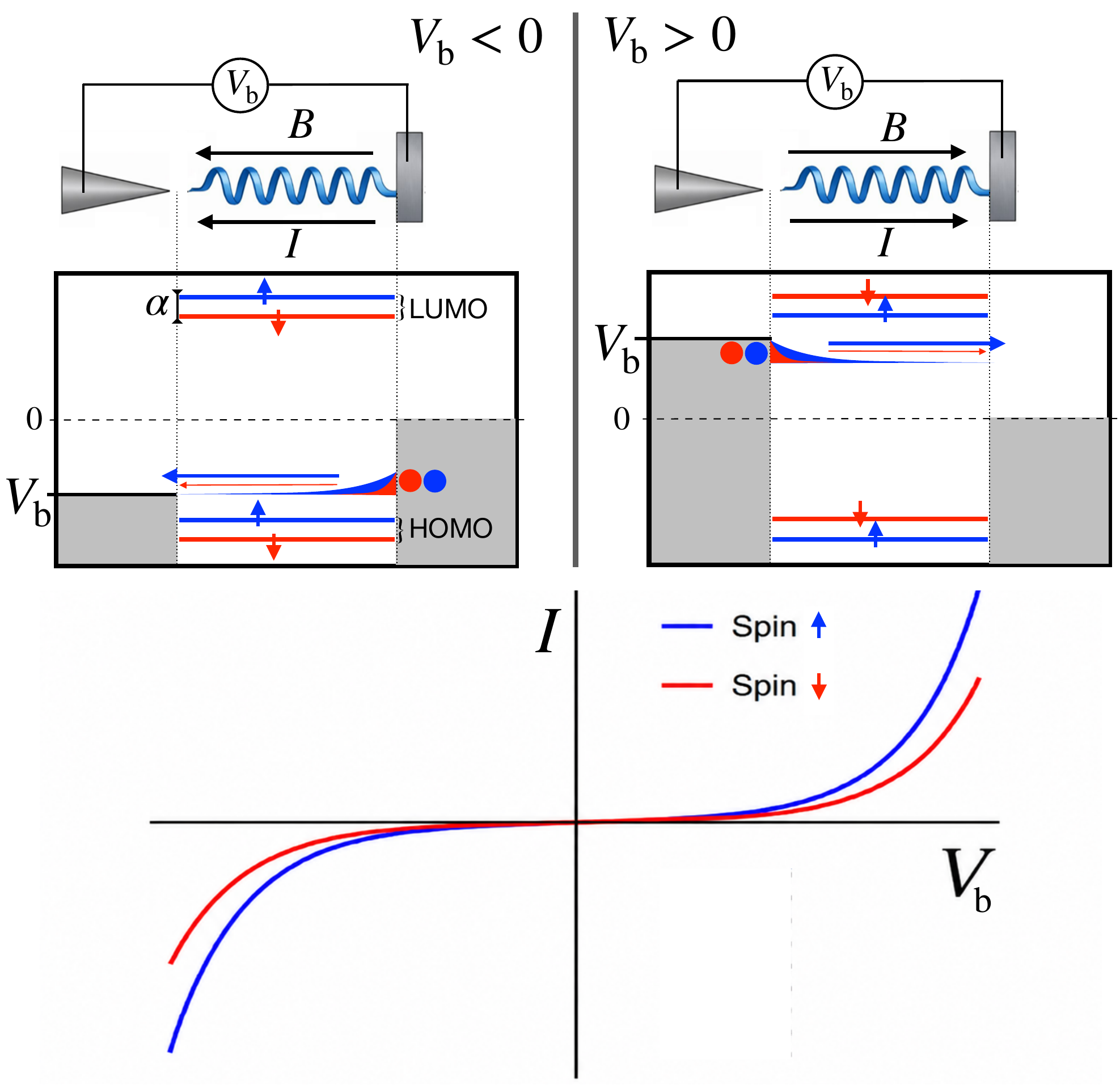}
\caption{\label{homo lumo} Schematic illustration of spin-dependent tunneling across a helical molecule with current-induced magnetic field $B$. For positive voltage bias (top right), the energy level of tunneling particles (filled circles) is closer to LUMO. Due to current-induced Zeeman splitting of LUMO, the barrier is lower for spin up (blue), which leads to a larger current for spin up than for spin down (see I-V curve below the figure). For negative voltage bias (top left), the charge current and thus the Zeeman splitting are opposite and the tunneling particles are closer to HOMO. The barrier is again lower for spin up. The resulting I-V curve (schematically reproduced at the bottom) has the characteristic CISS-like symmetry. Flipping the chirality of the molecule would flip the induced magnetic field and, hence, the roles of spin up and down.} 
\end{figure}

We now show that the considered model naturally explains the experimentally observed symmetry of the I-V curves, schematically shown in Fig.\ \ref{homo lumo}, namely the preserved sign of the SP under reversal of the current direction. 
We  consider the typical experimental setup of molecules placed on top of a metal substrate \cite{stm1, stm2} and accessed with a scanning-tunneling-microscopy tip. 
A voltage bias between the tip and the substrate is applied and the current flows between the substrate and the tip via the molecule. 
In such  asymmetric setups, the voltage drop is in general different at the two interfaces---one between molecule and the metal substrate, and one between molecule and the metallic tip. The exact voltage division might significantly depend on the experimental details such as the distance between the molecule and the tip \cite{stm3}. Here we adopt the most common situation in which the molecule couples more strongly to the metal substrate and the main voltage drop happens at the tip-molecule interface, in which case the potential of the metal substrate resides at the equilibrium value $E_F=0$, while the potential in the  tip, $ eV_\mathrm{b}$, changes with the applied voltage.
Therefore, in Eq.\ \eqref{current}, the Fermi-Dirac distribution functions assume the forms $f_L(E) =  (1+\exp[(E- eV_\mathrm{b})/k_{B}T])^{-1}$ and $f_R(E)=(1+\exp[E/k_{B}T])^{-1}$. Since we consider insulating molecules, the Fermi energy lies between the highest occupied molecular orbital (HOMO) and the lowest unoccupied molecular orbital (LUMO), as illustrated in Fig.~\ref{homo lumo}. 

The larger current is expected for that spin for which the tunneling barrier over the range of energies within the voltage window is smaller than for the other spin.  The smallest tunneling barrier is the energy difference between $eV_\mathrm{b}$ and the energy of the closest molecular level, which is either HOMO or LUMO, depending on the sign of the voltage bias.
These energy levels are spin-split due to the Zeeman term from the induced magnetic field, whose sign also depends on that of the voltage bias. Consequently, for a given spin projection, the barrier is lower or higher than for the opposite spin, independent of the sign of the bias voltage, as illustrated in Fig.~\ref{homo lumo}. The spin projection with the lower (higher) barrier thus has higher (lower) transmission and therefore larger (smaller) current at both positive and negative bias voltages, which explains the symmetry of the I-V curves.  It is interesting to note that the I-V curves can have the CISS symmetry also from  spin-dependent interactions other than via the self-induced magnetic field considered here. The crucial property of such an interaction is that it needs to be an odd function of the current, from which, together with time-reversal symmetry, follows that the tunneling barrier will be lower for the same spin for both current directions.
\begin{figure*}
\includegraphics[scale=0.55]{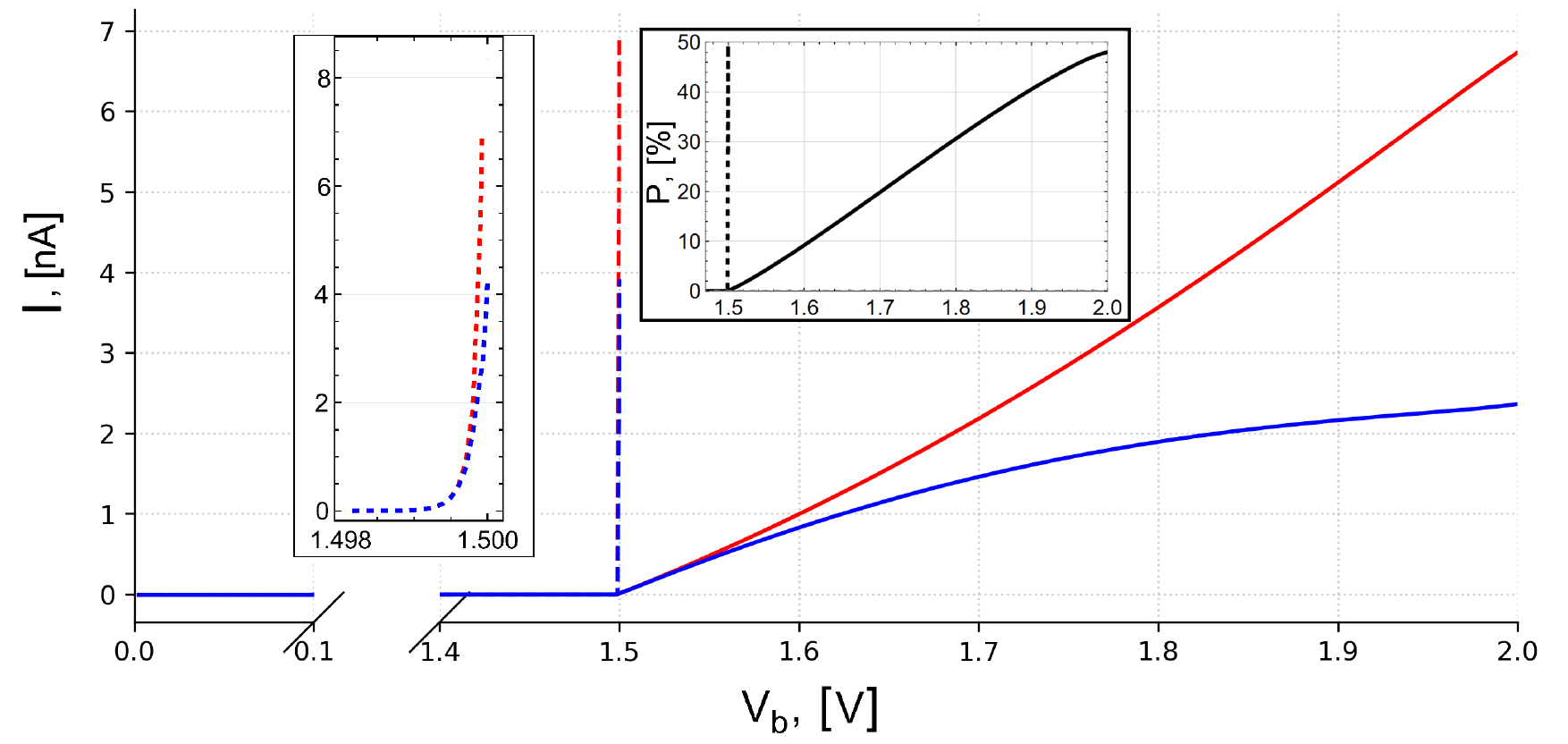}
\caption{\label{IVR} Current $I_+$ (blue) and $I_-$ (red) versus bias voltage $V_\mathrm{b}$ for $R = 0$ (dashed) and $R = 5.5 \times 10^7 \Omega$ (solid). The former shows  fast exponential growth near the barrier at $V_0=1.5$ eV with high spin splitting in a narrow region (see close-up in the left inset), while in the latter case the region of large SP becomes extended. The corresponding SP $P$ is shown in the right inset. Other parameters are $L = 12 $ nm, $m = 9 m_e$, where $m_e$ is the electron mass, $\alpha_0=10^{-4}$ eV, and $T = 0$ K. } \end{figure*}

The preserved sign of the SP under reversal of the voltage bias (I-V curve of  Fig.~\ref{homo lumo}) may appear to conflict with Onsager reciprocal relations, as has been pointed out in the literature \cite{ron, Yan_2024}. As they follow from microreversibility, they are valid only in the linear-response regime. The corresponding constraint on the conductance $G$ applies only in the zero-bias limit and can be written in the form
\begin{equation}
    G(I, \boldsymbol{B})\big|_{V_\mathrm{b} \rightarrow 0} = G(-I, -\boldsymbol{B})\big|_{V_\mathrm{b} \rightarrow 0}
\end{equation} 
which involves the reversal of all magnetic fields and currents of the system.
In our model, the current, and therefore the induced magnetic field responsible for the spin splitting, vanishes in the zero-bias limit, implying spin-independent conductance at $V_\mathrm{b}=0$. Instead, the SP in our model occurs at finite $V_\mathrm{b}$, i.e., in the non-linear response regime, so that there is no contradiction with Onsager relations.

\begin{figure}
\includegraphics[width=0.8\columnwidth]{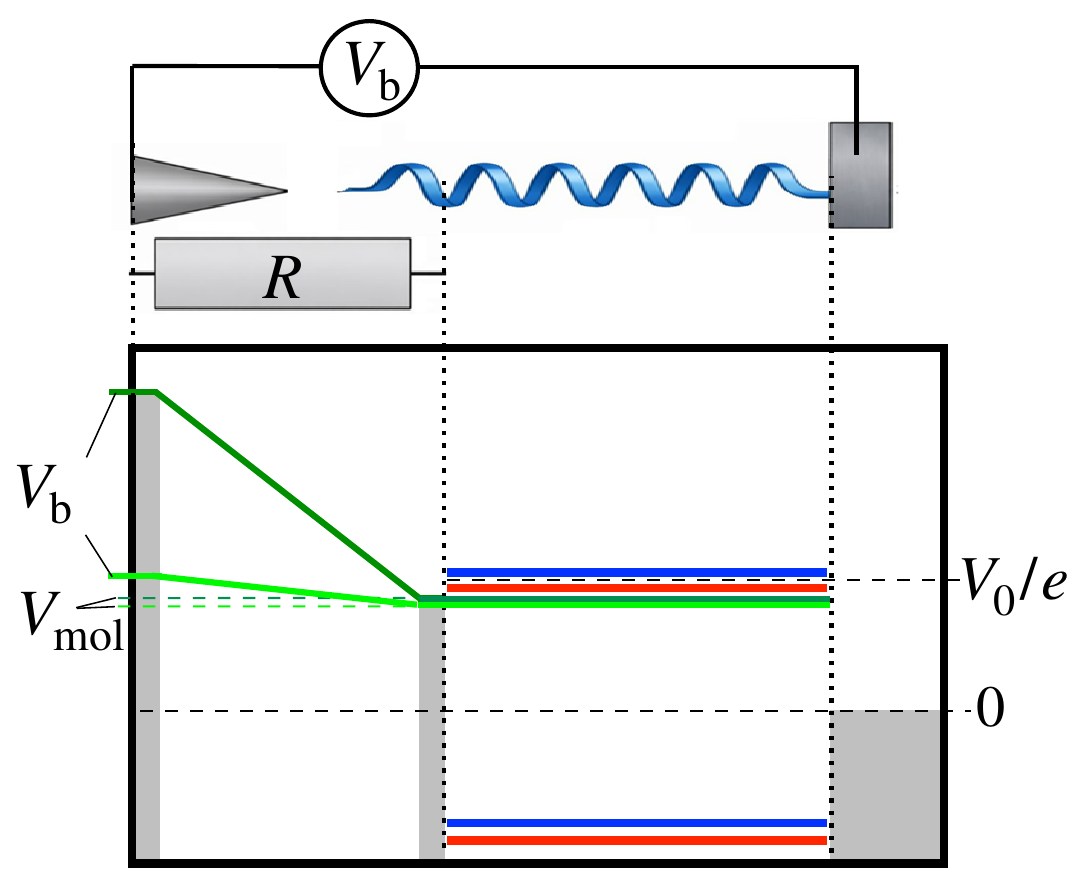}
\caption{\label{fig:resistor} Illustration of the effect of dissipation at the tip-molecule interface modeled by a resistor $R$. The potential $V_\mathrm{mol}$ stays close  below the energy levels of the molecule (red/blue lines) for an enhanced range of applied voltages $V_\mathrm{b}$ due to the current-dependent voltage drop at the resistor, $RI=V_\mathrm{b}-V_\mathrm{mol}$. Light and dark green lines illustrate potential profiles for two different values of $V_\mathrm{b}$.
} \end{figure}

\section{Spin polarization and the effect of dissipation}
\label{zeroT}

We now discuss the magnitude of the SP starting with the zero-temperature limit, where the Fermi-Dirac distributions are step-functions and the difference between them in Eq.~\eqref{current} is zero except for an energy window between $E=0$ and $E=eV_{\mathrm{b}}$, where it is unity. For simplicity (and since the symmetry of the I-V curves has been discussed above) we consider only positive $V_\mathrm{b}$. We compute the current using Eq.~\eqref{current} and find that the SP can be substantial in a small voltage bias range (Fig.~\ref{IVR}, dashed lines). This can be understood from the parameter dependence of the spin-split transmission in Eq.\ \eqref{eq:tsplit},  which gives a sizeable splitting only for energies within the distance $\alpha$ from the barrier, $|V_0-E|\lesssim\alpha\sim 10^{-4}$ eV. 

The narrow voltage range of sizeable SP can, however, become substantially extended if the interface between the molecule and the tip is dissipative. In the considered setups, the dissipation occurs naturally, e.g., due to coupling of the tunneling particle to vibrational modes at the loose end of the molecules \footnote{Relevant experiments showing large spin polarization are typically performed on a film of molecules, where inter-molecular interactions may provide additional dissipation channels.}. We do not need to assume that the dissipation is spin-dependent. Any small spin dependence would only weakly modify our results, since the main SP comes from the spin-dependent tunneling probability, while the role of dissipation is merely to extend the voltage window (as will become clear below). To account for the dissipation in our model, we consider a (spin-independent) resistance $R$ between the tip-electrode and the molecule as illustrated in Fig.\ \ref{fig:resistor}.
A part of the applied voltage $V_\mathrm{b}$ will drop at the resistance, so that the voltage throughout most of the molecule has the value
\begin{equation}
\label{vmol}
    V_{\mathrm{mol}} = {V_{\mathrm{b}}} - IR.
\end{equation}
 In the calculation of the total current, $I=I_++I_-$, we thus replace $V_\mathrm{b}\to V_{\mathrm{mol}}$ in Eq.~\eqref{current}, to be solved self-consistently.
 To gain an intuition for the effect of the resistor $R$ it is useful to consider the tunnel current in the WKB approximation (not applied to the numerical results below), reading 
 \begin{equation}
     I_\sigma \approx I_0\, e^{-2\kappa L\sqrt{(V_\sigma-eV_\mathrm{mol})/\left[\mbox{eV}\right]}},\ \ \ \ \label{iapprox}
 \end{equation}
 where 
 \begin{equation}
     I_0=  \frac{2}{(\kappa L)^2}\frac{e}{h} \left[\mbox{eV}\right],\ \ \ \kappa L =\frac{\sqrt{2m[\mbox{eV}]}}{\hbar}L
     \label{iwkb}
 \end{equation}
 are the maximal current $I_0$, reached at $eV_\mathrm{mol}=V_\sigma$, and the dimensionless length of the molecule $\kappa L$, respectively. For a typical molecular length  $L\sim 10$ nm and an enhanced mass, the dimensionless length can be on the order of $\kappa L\sim 100$, hence $I_0\sim1$ nA, which is the characteristic current magnitude in relevant experiments.  

From Eq.\ \eqref{iapprox} we note that tunneling occurs as long as $ V_\mathrm{b}<V_\sigma/e+IR$. Since $I$ is a monotonic function of the voltage and the maximal current is $I_0$ we obtain
\begin{equation}
\label{extension}
    V_\mathrm{b} \leq \min_\sigma(V_\sigma/e) + I_0  R
\end{equation}
expressing the upper bound for the bias voltage of the tunneling regime. A resistance of order $R\sim[1\text{V}]/I_0$ thus extends the bias-voltage range by several volts above the energy of the molecular orbital (as illustrated in Fig.\ \ref{fig:resistor}). We will now show that the current and SP in this extended range are large, i.e., current on the order of $I_0$ and  SP on the order of $100\%$. 
First note that for a bias voltage in the range $V_\sigma/e<V_\mathrm{b}<V_\sigma/e + I_0  R$, say at $V_\mathrm{b}=V_\sigma/e+I_0  R/2$,  the total current is on the order of $I_0$, since the voltage drop at the resistor must be at least $RI_0/2$. The resistor thus effectively inflates the voltage range for which the potential at the molecule is closely below the barrier. Consequently, the SP of the current is large, as can  be estimated using the WKB approximation for the current \eqref{iapprox}. Since $ I \sim I_0$ and $L\kappa \sim 100$, we can approximate $eV_\mathrm{mol}\approx V_0-\alpha(I)$ and obtain
\begin{equation}
   P \sim \tanh \big(L\kappa\sqrt{2\alpha_0/[\mbox{eV}]}\big).
   \label{papprox}
\end{equation}
For realistic molecule lengths and current-induced Zeeman splitting, $L\kappa \sim 100$ and $\alpha_0\sim 10^{-4}$ eV, respectively, the SP can thus be on the order of 100\% within a finite voltage range.

We confirm this analysis with full, self-consistent numerical calculations (without WKB approximation), whose results are shown in Fig.~\ref{IVR}. While for $R=0$ a large SP is reached only in a vanishing voltage range close to resonance with the molecular orbital, this range becomes finite in the presence of a resistor. 

\begin{figure}
\includegraphics[scale=0.3]{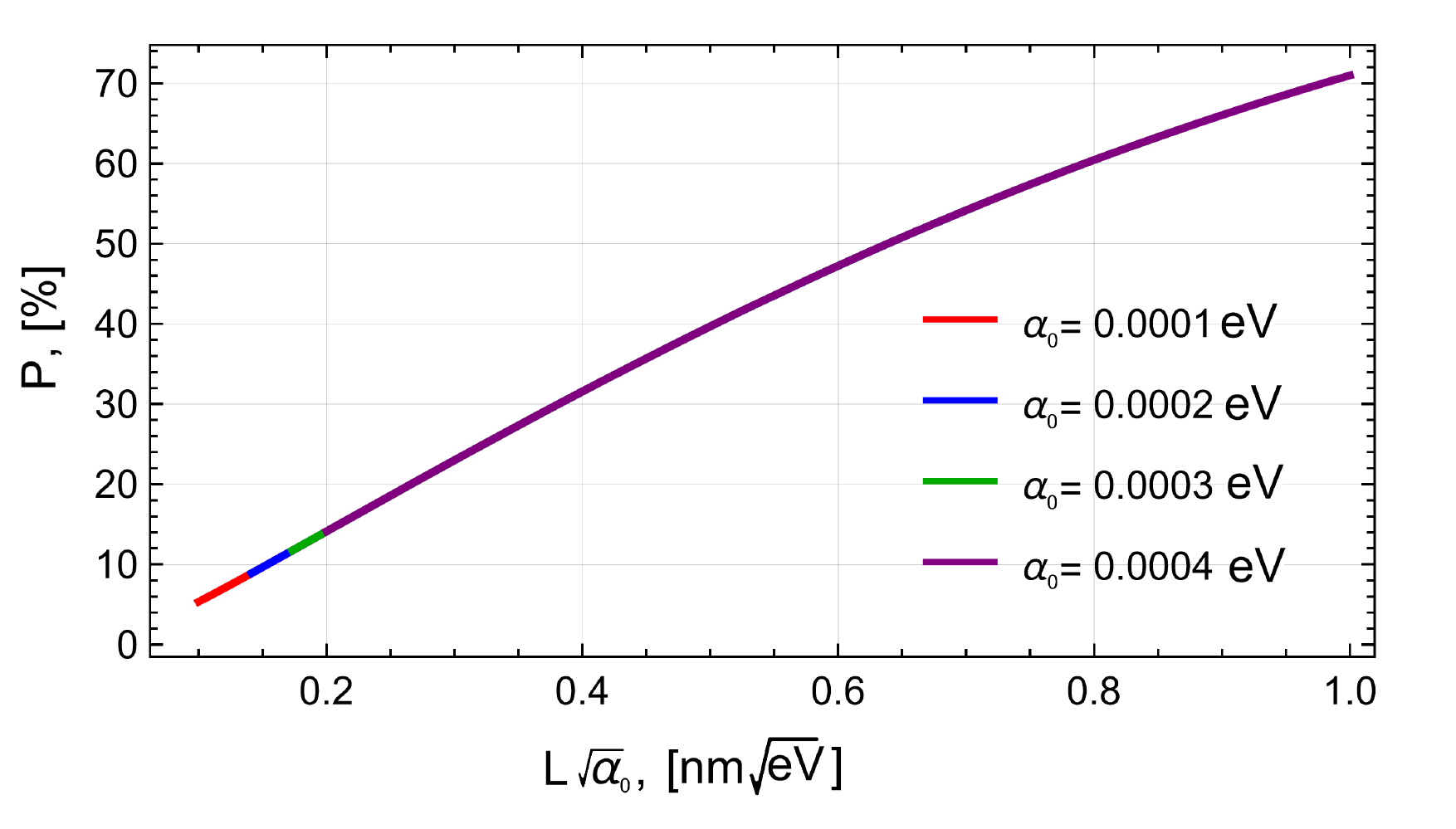}
\caption{\label{alpha}SP plotted as a function of the length of the molecule against $L \, \sqrt{\alpha_0}$. The parameters for these plots are $T = 0$ K, $V_0 = 1.5$ eV, $R = 10^5\Omega$, $V_\mathrm{b} = V_0 + (3/4) I_0R$.}
\end{figure}

In Fig.\ \ref{alpha} we plot the SP for different $\alpha_0$ values against the scaled axis $L  \sqrt{\alpha_0}$. All curves with different $\alpha_0$ fall onto the same line, which confirms the scaling of the SP with the system length $L$ as expressed in Eq.~\eqref{papprox}. Our results thus show that the smallness of the spin-splitting perturbation $\alpha_0$ can be compensated by the length $L$ of the molecule and the effective mass $m$ (through $\kappa$), which can be enhanced due to coupling to phonons \cite{Klein2023}.
\begin{figure*}
\includegraphics[scale=0.6]{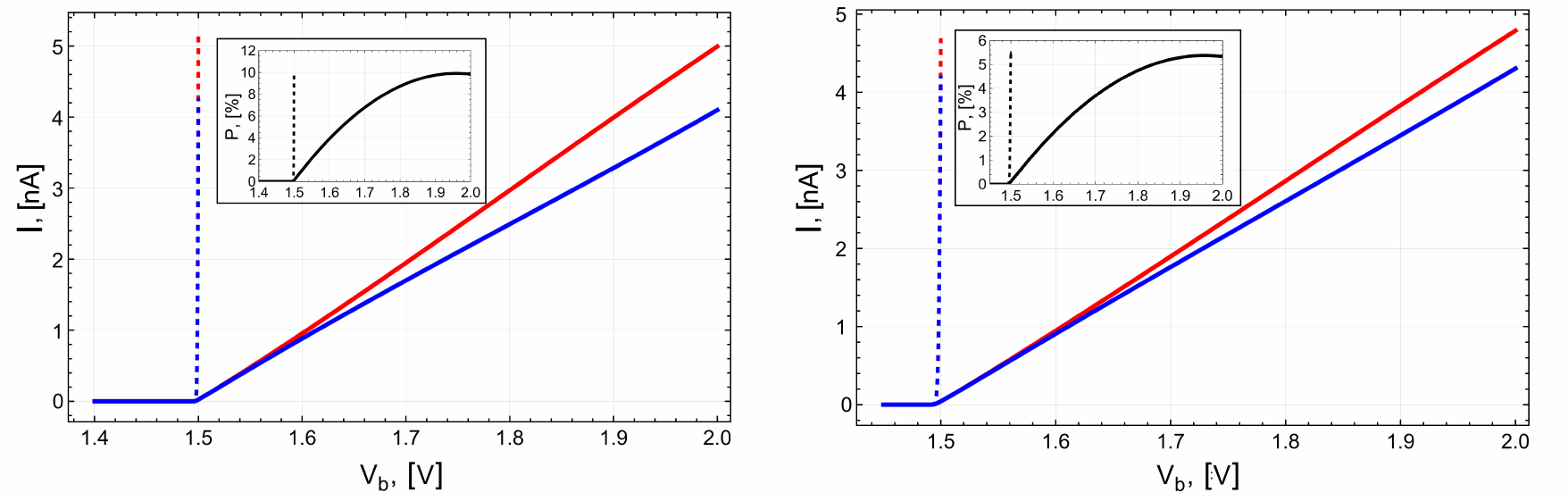}
\caption{\label{T=5} Current $I_+$ (blue) and $I_-$ (red) as a function of bias voltage $V_\mathrm{b}$ at temperature $T=5$ K (left) and $T=10$ K (right). Dashed curves correspond to $R=0$ and solid lines to $R = 5.5 \times 10^7\Omega$. Insets show the corresponding SP. Other parameters are
$V_0 = 1.5$ eV, $L = 12$ nm, and the effective mass $m = 9 m_e$.}
\end{figure*}
\section{Spin polarization at finite temperature}
\label{5}
We now extend our analysis to the finite-temperature case.  In our model, the finite temperature leads to thermal broadening of the distribution functions, so that even for bias potentials below the tunneling barrier, thermally excited electrons can gain enough energy to transmit above the barrier. 
The integrand of Eq.~\eqref{current} is the product of the transmission and distribution functions, which both are  exponentially decaying. It has two peaks, at approximately the bias potential $E=eV_\mathrm{b}$ and the potential of the molecular orbital $E=V_\sigma$, which correspond to tunneling particles and thermally excited particles, respectively. The height of the former is determined by the tunneling transmission, $\sim\exp[-L\kappa\sqrt{(V_\sigma-eV_\mathrm{b})/[\mbox{eV}]}\big]$, and that of the latter by the decaying tail of the Fermi-Dirac distribution, $\sim \exp[-(V_\sigma-eV_\mathrm{b})/k_BT]$.
While for energies below the barrier, the transmission function \eqref{T}  exponentially decays with the length of the molecule $L$, above the barrier the transmission is of order one with an oscillatory behavior. The large SP that can be achieved in the tunneling regime cannot be achieved above the barrier, where the difference of transmissions for the two spins oscillates as a function of energy.

For voltages close to the barrier significant contributions from thermally excited particles thus suppress the SP. This is seen in Fig.\ \ref{T=5}, where we plot the I-V curves for the same parameters as in Fig.~\ref{IVR} but at finite temperature. As expected, the SP decreases with increasing temperature as soon as $k_BT\gtrsim \alpha_0$.
\section{Discussion}
\label{summary}
In the broader context of existing theoretical and experimental studies of CISS, the experimentally reported values of SP range from a few percent up to values approaching 100\%, depending on the experimental setup, molecules, and the measurement protocol. In contrast, microscopic theoretical models based on realistic spin-orbit-coupling strength typically yield much smaller SPs even when including possible enhancement mechanisms such as those due to vibrations. 
According to our model, the large SP up to 100\% can stem from the exponential sensitivity of the transmission probability to small spin-dependent perturbations in the quantum-tunneling regime. We exemplified such perturbations using the model of Zeeman interaction with a current-induced magnetic field, but we expect that other perturbations stemming from electron correlations can play a similar role.
Furthermore, our tunneling model, where due to the asymmetry of the tip-molecule-substrate junction the tunneling goes via HOMO or LUMO, depending on the sign of the applied voltage, naturally leads to a CISS-like symmetry of I-V curves. Thereby, Onsager's relations are not violated since the SP in this mechanism occurs only in the non-linear response regime.

An important ingredient in this scenario is dissipation at the tip-molecule interface, which we modeled by a generic, spin-independent resistor. The dissipation in our model is not a source of spin polarization; its effect is to  inflate the bias-voltage range, over which the potential throughout the main part of the molecule stays close to the tunneling barrier and the large SP occurs. We suspect that such dissipation is enhanced in setups of molecular films with dangling molecule ends due to enhanced vibrational degrees of freedom of the film. This possibly explains why a large SP has been observed in experimental setups where the junction is made of a metallic tip approaching a film of molecules protruding from a metal substrate, but no SP was observed in single-molecule break junctions \cite{Li2025}, where the junction consists of a single molecule rigidly coupled to two electrodes. 

Within our model, temperature introduces a competition between under-the-barrier tunneling and thermally activated over-the-barrier transport, which leads to suppression of SP with temperature. The experimentally observed temperature dependence of SP remains controversial. While some experiments report robust SP at room temperature \cite{ciss_dna, temp_review}, there are also indications that the effect can become more pronounced at lower temperatures, depending on the transport regime and molecular environment \cite{low1, low2}. In this context, our model may provide a useful perspective for interpretation and design of temperature-dependent measurements. Further insights from the theory side may be provided by an extension of our model to include inelastic tunneling. 
\begin{acknowledgments}
We thank Dirk Morr, Robert Bittl, Vladimiro Mujica, and Ferdinand Evers for useful discussions. This work was supported by the Deutsche Forschungsgemeinschaft (DFG, German Research Foundation) - Project Number 277101999 - CRC-TR 183 (project A02), the Emmy Noether program - Project Number 506208038 -, and Cluster of Excellence EXC 3112 Center for Chiral Electronics.
\end{acknowledgments}

\bibliography{library}
\end{document}